\documentclass[11pt]{article}
\usepackage{graphicx}
\usepackage[margin=1.25in]{geometry}
\usepackage[usenames,dvipsnames]{color}
\usepackage{url}
\usepackage[colorlinks = true,
            linkcolor = blue,
            urlcolor  = blue,
            citecolor = blue,
            anchorcolor = blue]{hyperref}

\newcommand\pubnumber{PUBNUMBER }
\newcommand\pubdate{\today}

\def\Title#1{\begin{center} {\LARGE #1 } \end{center}}
\def\Author#1{\begin{center}{ \sc #1} \end{center}}
\def\Address#1{\begin{center}{ \it #1} \end{center}}

\newcommand\pubblock{\rightline{\begin{tabular}{l} \pubnumber\\
         \pubdate \end{tabular}}}
\newenvironment{Abstract}{\begin{quotation} \begin{center}
                       ABSTRACT
     \end{center}\bigskip  }{\end{quotation}}

\newcommand\snowmass{\begin{center}\rule[-0.2in]{\hsize}{0.01in}\\\rule{\hsize}{0.01in}\\
\vskip 0.1in Submitted to the  Proceedings of the US Community Study\\ 
on the Future of Particle Physics (Snowmass 2021)\\ 
\rule{\hsize}{0.01in}\\\rule[+0.2in]{\hsize}{0.01in} \end{center}}

\begin{document}

\pubblock

\Title{Evolution of HEP Processing Frameworks}

\bigskip 

\Author{Christopher D. Jones and Kyle Knoepfel}
\Address{Fermi National Accelerator Laboratory, Batavia, IL 60510, USA}
\medskip
\Author{Paolo Calafiura, Charles Leggett, and Vakhtang Tsulaia}
\Address{Lawrence Berkeley National Laboratory, Berkeley, CA 94720, USA}
\medskip

\begin{Abstract}
\noindent HEP data-processing software must support the disparate physics needs of many experiments. For both collider and neutrino environments, HEP experiments typically use data-processing frameworks to manage the computational complexities of their large-scale data processing needs.

Data-processing frameworks are being faced with new challenges this decade. The computing landscape has changed from the past three decades of homogeneous single core x86 batch jobs running on grid sites.  Frameworks must now work on a heterogeneous mixture of different platforms: multi-core machines, different CPU architectures, and computational accelerators; and different computing sites: grid, cloud and high-performance computing.

We describe these challenges in more detail and how frameworks may confront them.  Given their historic success, frameworks will continue to be critical software systems that enable HEP experiments to meet their computing needs. Frameworks have weathered computing revolutions in the past; they will do so again with support from the HEP community.
\end{Abstract}

\snowmass

\def\thefootnote{\fnsymbol{footnote}}
\setcounter{footnote}{0}

\section{Executive Summary}

The recent changes to computational hardware and to the kinds of sites hosting that hardware in the past decade calls for renewed research and development into HEP data-processing frameworks in order fully exploit these resources. Computational resources of the previous decades were  dominated by the world-wide research grid composed of commodity x86 CPUs running the Linux operating system. Experiments just needed to target that hardware with single-threaded applications in order to meet their computational needs. The advent of multi-core systems, the prevalence of non-CPU based computational hardware, and the growth in use of non-grid computational sites all require new research and development at the framework level. Multi-core systems necessitate exposing and exploiting software concurrency. New kinds of computational hardware require new ways to schedule work in the application and potentially new third party library interfaces to accommodate. High Performance Computing and commercial cloud sites bring unique requirements and possibilities which will shape additional aspects of frameworks.

Modifying experiments software frameworks to support multi-processing and multi-threading event processing models significantly improves utilization of shared system resources on modern many-core architectures, and also increases the overall event processing throughput. Current software frameworks of many HENP experiments are capable of concurrently processing multiple events either within a single multi-threaded application or by coordinating event processing across multiple processes. To further optimize resource utilization of certain workloads, some experiments have implemented hybrid multiprocess/multithreaded processing models. The first results obtained in this area are encouraging, and further research and development should be made into exploring the possibility of a generalized hybrid framework that can combine the strengths of the two parallel processing models.

Computational accelerators (e.g., GPUs, FPGAs) are becoming increasingly prevalent in most new computing facilities. To port HEP applications to these architectures, significant technical challenges must be addressed, such as efficiently scheduling work on both accelerators and CPUs, accommodating the various programming models of accelerators and CPUs, and adjusting experiment data models to be more compatible with accelerator expectations.

While the world-wide computing grid continues to be an important computational resource for HEP, it is expected that in the future, experiments will rely more on HPC sites and commercial cloud-based resources. Each of these resource types requires some specialization by the frameworks to achieve optimal utilization.  Cloud-computing sites also bring a new set of requirements for frameworks, as such sites are likely to be used on an opportunistic basis. This may require that the event processing jobs must be able to terminate on a short notice and preserve all work done so far, or they must periodically save the event processing results.

Framework developers will continue to rely on the firm support they have receive from HEP experiments as they confront the computing challenges of the next decade.

\section{Introduction}

Although the term \textit{framework} can be used in multiple ways, we use it to denote a software system that HEP experiments use to simulate or reconstruct detector data with the goal of extracting physics quantities of interest---or \textit{data products}.  HEP frameworks provide non-trivial data-processing functionality that includes:
\begin{itemize}
    \item The ability to schedule nearly arbitrary user-defined algorithms for a program's execution.
    \item A provenance subsystem that records the processing history of data products.
    \item Organizing data into physically meaningful datasets or periods of time.
    \item Managing iteration over events and other data groupings.
\end{itemize}
Because physicists rely on these features as part of the analysis process, frameworks have become ``an essential part of HEP experiments’ software stacks"~\cite{calafiura2019hep}.

The ways in which frameworks have provided these capabilities, however, have changed over time.  Not only do newer computing technologies necessitate rethinking long-established programming paradigms, but changes in the amount of data to be processed---and the hardware on which the processing occurs---also warrant reassessing the \textit{status quo} of HEP computing.

In section~\ref{sec:many-core} of this document we discuss various models of parallel event processing (e.g., multithreaded, multiprocess, hybrid) adopted by the experiments in order to efficiently utilize computing resources of modern many-core nodes. Section~\ref{sec:arch} describes the challenges posed to experiments' software frameworks by the need to support wide range of available CPU architectures. In section~\ref{sec:accelerators} we describe the differences between heterogeneous (e.g., GPU, FPGA) and traditional CPU architectures, we also discuss various GPU programming models available today on the market, and the challenges related to offloading computations from the frameworks to accelerators. Finally, section~\ref{sec:sites} describes three main categories of computational sites---grid, HPCs and commercial clouds---and discusses some specifics of running experiments' data-processing applications on these sites.

\section{Many-Core}
\label{sec:many-core}

For decades, it was sufficient for an experiment's data-processing framework to execute on only one CPU core. Computing clusters and large-scale batch systems were deployed and have achieved impressive data-processing throughput using only commodity CPU hardware.  HEP experiments also took advantage of the ``perfect parallelization'' of event-processing made possible by the statistical independence of each recorded event. This parallelization was achieved by having many independent single-threaded processes each independently processing their own group of events.  However, as CPU clock rates have plateaued, computing nodes have changed to including multiple integrated CPU cores, performing computations in parallel to keep up with Dennard scaling~\cite{dennard}. Efficiently utilizing nodes with large numbers of cores can be challenging for traditional computing models where each job runs independently on a single core. This can lead to very large memory footprints, and inefficient use of system resources. Modifying these frameworks to support multiprocessing and multithreading on many-core architectures can significantly improve utilization of shared system resources.  In addition, a single multi-core job can process a larger number of events within the same amount of time, with all the events being written to the same file. The resultant larger files can then either limit or completely eliminate the need to later merge output files together in order to generate files of sufficient size to be efficiently stored to archival media, such as tape. In addition, the total number of multi-core jobs needed to complete a large workflow is substantially less than the number of single-threaded jobs needed to do the same workflow. This decrease in total number of jobs to manage can significantly decrease the pressure placed on the workflow management system.

Within the last 10 years, many High Energy and Nuclear Physics experiments have gained the ability to use multiple cores within a single job on a machine when doing production work (such experiments include ATLAS and LHCb through the Gaudi framework~\cite{Clemencic_2015}, CMS~\cite{Jones_2015}, ALICE, Mu2e through the \textit{art} framework~\cite{Green_2012}, GlueX through the JANA2 framework~\cite{jana2_chep19}, and Jefferson Laboratory's CLARA framework~\cite{CLARA}). These experiments' frameworks are now able to concurrently process multiple events within one multithreaded application or by coordination across multiple processes.

\subsection{Multithreaded Processing Model}

Multithreading sees a single process running multiple threads of execution where all threads share the same memory space but each thread performs its own operations. Communication across threads is extremely fast (as it is just read/write to shared memory) but unintended sharing (known as race conditions) can be very difficult to diagnose and fix. The memory sharing across threads allows a single, multithreaded program to consume less CPU memory than the equivalent number of concurrently running, single-threaded applications. The memory savings were a major reason for ATLAS and CMS to move to a multithreaded framework. multithreading technologies make it easy to schedule tasks for concurrent execution on a machine's available CPU cores. By splitting the work into a large number of small tasks, the concurrency opportunities increase even further. 

\subsection{Multiprocess Processing Model}

In contrast to multithreading, multiprocess involves executing multiple intercommunicating processes on one or more compute nodes. These processes each have their own mutable memory space which avoids the unintended sharing difficulties of multithreading. It is possible to avoid redundant copies of immutable data (e.g. detector geometry descriptions) either by sharing the memory via the operating system's ``copy-on-write'' feature, or by partitioning the tasks assigned to each process such that all tasks using the same immutable data are in the same process. Avoiding such redundancies can greatly lower overall memory consumption. As just mentioned, the individual processes are usually assigned a specific set of tasks they will perform. Communication across processes is, in general, slower than within a process (as data transferred between processes often needs to be packed, transferred, and then unpacked). This tends to require that, in order to amortize communication time, the tasks performed by each process do more work and run longer than tasks in a multithreaded framework. Given the restriction on tasks done by each process, it can be more challenging to efficiently use the available CPU cores when using multiprocess implementations as compared to multithreading.

\subsection{Hybrid Processing Models}

Just as it is possible to have multithreaded processes be part of a multiprocess based system, it is also possible for a single, multithreaded application to control multiple external processes. This is used by the CMS experiment to run concurrent single-threaded instances of Monte Carlo generators where the generators are not safe to run multithreaded. In this case, the main threaded process has a task associated with each generator instance and when the main application wants to run that generator, the associated task pauses the thread in the main application and then permits the generator application to ``take over'' that core for its own execution. Once the generator finishes processing the event, the generator process pauses and the main application's task wakes up, transfers the data between processes and then returns control to the main application. In this way, the external process is essentially scheduled just like a task within the main application.

Another approach to the hybrid processing model is to implement a multiprocess architecture where each event-processing unit is a multithreaded process. The usage of this approach is being considered by the ATLAS experiment for some workloads with high mutable memory footprint (hence not suitable for running in pure multiprocess mode) that have not demonstrated good scaling of event-processing throughput due to lock contention among threads. The exact configuration of threads and processes can be fine-tuned to specific hardware and job requirements.

Further research and development should be made into exploring the possibility of a generalized hybrid multiprocess/multithreaded HEP framework system that can combine the strengths of the two processing models.

\subsection{Further Considerations}

Many applications now also support concurrent processing of tasks at the sub-event level. For most HEP experiments, this level of concurrency is more limited than event-level parallelism as the number of tasks tends to top out around 10,000 independent tasks (e.g. the number of charged-particle tracks in an extreme collision) versus the 100s of billions of independent events that are typically processed by the LHC experiments. In addition, the time required to execute intra-event tasks in parallel tends to be a small fraction of the total time needed to process an event; by contrast, concurrently processing separate events naturally fills an event's entire processing time. For neutrino experiments such as DUNE, however, the trade-offs between inter- and intra-event parallelism may be different as the number of events to process at a given time may be less than the available cores. In addition, trying to process as many concurrent events as cores on a machine may not be possible if the mutable memory needed per event exceeds the available memory per core on a standard machine. Given the different kinds of exploitable concurrency, it may therefore be necessary to develop and maintain different frameworks based on characteristics of the data being collected and the tasks to be performed.

As frameworks apply more cores to processing data from the same source (e.g. a file on disk) and writing data to the same storage (e.g. another file on disk), the serial nature of traditional I/O systems becomes a limit to the concurrency scaling. CMS has already found that attempting to scale beyond 8 threads per job while doing reconstruction leads to an overall CPU efficiency of less than 80\% because of the need to sequentially call into ROOT for I/O. A component of the US DOE CCE project~\cite{CCE} is researching ways to allow greater concurrency while doing I/O but additional research will probably needed in this area in order to exploit a greater number of cores.

\section{CPU Architectures}
\label{sec:arch}

The dominance of commodity Intel x86 compatible hardware as the major source of HEP computing resources appears to be coming to an end. As more HEP computational resources come from high-performance computing (HPC) sites, the types of CPU architectures frameworks must support will grow.

One of the driving factors in the design of HPC sites is the computational energy efficiency---i.e. the energy utilization per floating-point operation. The use of traditional x86-based homogeneous CPU designs for leadership class HPC facilities has proved unfeasible due to their electrical demands. Instead, increasingly large fractions of silicon are being devoted to more energy-efficient computational accelerators such as GPUs, as well as a shift towards lower-power traditional CPUs such as ARM-based chips. Examples of these are Frontier at ORNL, where more than 90\% of the computing power comes from AMD GPUs, and Fugaku in Japan, the current leader of the TOP500 supercomputers, which is based on a Fujitsu AF64 ARM 8.2A core. Manufacturers have also started offering less monolithic chips, where a single CPU is composed of multiple cores of different power and performance, or where one or more traditional CPUs are bound together with a GPU or FPGA accelerator in a chiplet array, offering much higher bandwidth between components than with a traditional add-in accelerator card.  

\begin{figure}
    \centering
    \includegraphics[width=0.95\textwidth]{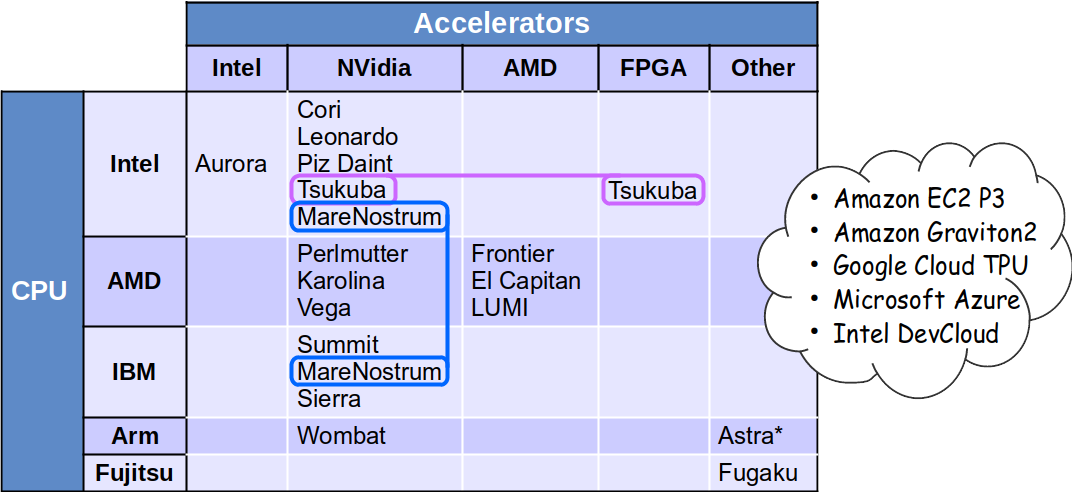}%
    \caption{Current HPC landscape.}
    \label{fig:HPCs}
\end{figure}

The validation of physics results across these different architectures may require software experts to assist in proper handling of portable floating-point evaluations to guarantee identical results. Experiments may, otherwise, have to give up on bit-wise reproducibility of results and favor statistical equivalency instead. Either way, the frameworks will need to record information about what architectures were used for each generated data product to assist in debugging code.

In order to get the best usage of different CPU architectures, we may be required to compile our code either with specific architectural compiler directives or with specific compilers. This approach will likely lead to many variants of an experiment's code base which will have to be managed and distributed. Presently, one complete build of CMSSW (including all third party libraries) requires approximately 100GB of storage space. Having multiples of these for each software architecture, and then within a given architecture multiple variants for each sub-hardware configuration, is likely to be unsustainable.  To mitigate this problem, development effort is needed to explore viable solutions, perhaps including:

\begin{itemize}
    \item \emph{Fat binaries}, where one build contains different machine code versions of the same function, and the language or operating system automatically uses the correct version at run-time.
    \item \emph{Selective optimization}, where the developer identifies which framework components would benefit the most from architecture-specific builds and the framework would dynamically load the selected components built for specific hardware.
    \item \emph{On-site builds}, where the code is built directly on the site that will run it. Traditional build mechanisms have made such an idea extremely difficult. With the advent of better build tools, such as Spack~\cite{Gamblin_The_Spack_Package_2015}, this approach becomes more feasible, although the size required for each build may still be very large.
\end{itemize}
For each of the above options, the size and number of different builds that may be needed to support all the workflows planned to run at a site, and the validation of the physics results for those builds are still challenges that would need to be confronted.

\section{Computational Accelerators}
\label{sec:accelerators}

Computational accelerators, such as GPUs, and to a lesser extent FPGAs, are becoming increasingly prevalent in most new computing facilities (see Figure~\ref{fig:HPCs}) due to their reduced energy consumption. There exist major challenges, however, in porting current HEP workflows to these architectures. Traditional programming paradigms commonly found in HEP programs do not map well onto GPU architectures.

GPUs and FPGAs are fundamentally different from traditional CPU architectures. Although affording massive parallelism, the memory access patterns and high latency of GPUs require a fundamentally different programming model than that of CPUs. Other accelerators such as TPUs are optimized for specific tasks, such as matrix manipulations, that are heavily used in machine learning workflows. While these can enormously accelerate the computations for which they were designed, they are very inefficient as a general purpose processor. Field programmable devices such as FPGAs and ASICS are also very efficient for the specific algorithms for which they are designed, but require time consuming reprogramming to reconfigure for a different task. 

\begin{figure}
    \centering
    \includegraphics[width=0.95\textwidth]{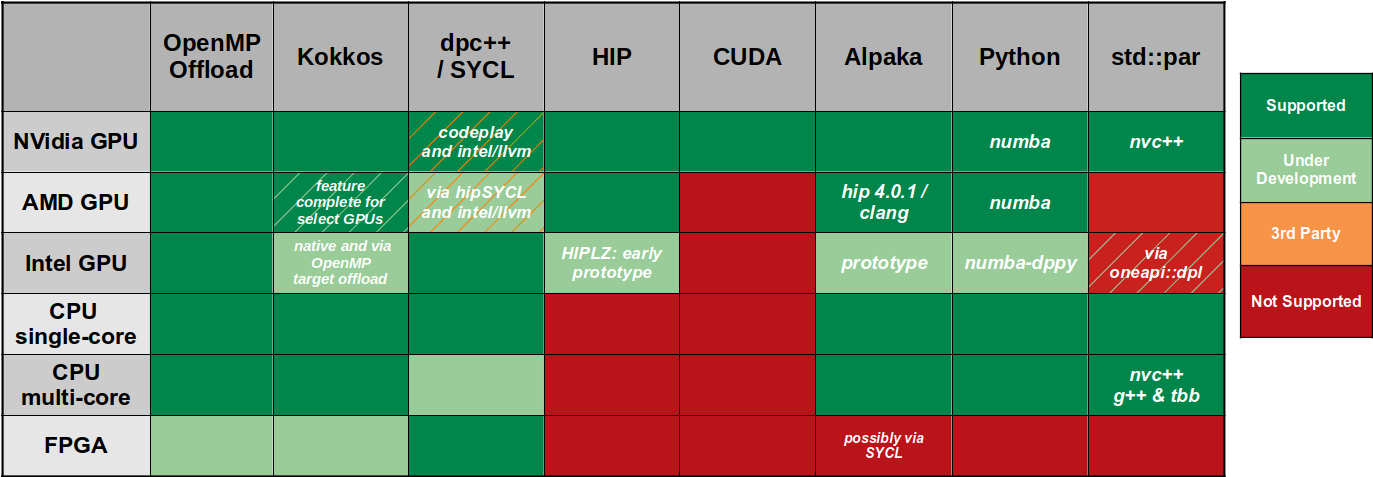}%
    \caption{Portability solutions for heterogeneous architectures.}
    \label{fig:portability}
\end{figure}

The situation is further complicated by the fact that the three major producers of GPUs, NVIDIA, AMD and Intel, each promote their own programming model and development environment, which tend to be incompatible with each other. This use of device specific languages requires the re-coding or translation of source code for the appropriate device, which can be very time consuming given the large code base that HEP experiments have developed. The human cost of developing, validating, and maintaining multiple versions of source code for each architecture is too burdensome for any experiment to currently contemplate. A number of portability layers are available (see Figure~\ref{fig:portability}), such as Kokkos~\cite{kokkos}, Alpaka~\cite{Alpaka}, OpenMP~\cite{OpenMP}, and SYCL~\cite{SYCL}, and standards are slowly emerging that will eventually percolate into the C++ language standard. However, even with a portability layer that can target multiple different hardware backends from the same source code, different libraries, compiler options, and sometimes even different compilers are currently needed to generate the binaries, thus increasing build complexity and code distribution overheads.

Integrating computational accelerators into traditional frameworks adds an extra level of complexity. The optimal technique of executing code on accelerators is usually via asynchronous offloading, where once a task has been sent to the accelerator by the framework executing on the CPU, it does not need to wait idly for the results, and can instead execute other useful work. Once the accelerator has finished its task, the framework must then be informed so that the results can be retrieved and downstream execution can continue. This asynchronous model can often require significant modifications to existing frameworks that have been developed for traditional CPUs. Furthermore, the amount of ``work'' that can be offloaded to an accelerator from a single event is often rather small, leading to inefficient use of the device. In order to maximize accelerator performance, data may need to be batched between multiple events, which is a significant complication for frameworks that normally consider each event to be completely independent. 

Many modern frameworks have evolved complex event data models (EDMs) that make extensive use of object-oriented features such as multiple levels of polymorphic inheritance. These types of data structures cannot be used efficiently on accelerators, which prefer flat structures where threads access sequential or regularly stepped memory locations. This has caused the need to reformulate data objects that can be accessed on accelerators, either by explicitly changing the EDM, which may involve modifying large amounts of associated code, or by converting them at run-time as they are transferred to and from the device, which can impose a significant performance penalty. Further development is needed in this area to find the best trade-offs between performance and usability.

\section{Computational Sites}
\label{sec:sites}

For several decades HEP computing has relied on a world-wide computational grid to do the bulk of computing for the experiments. The grid has been composed of machines running compatible versions of Linux on x86-compatible hardware. This large-scale homogeneity has made it easy to compile an experiment's code once and then use it everywhere. The grid will continue to be an important computational resource for HEP, but in the future, experiments will rely more on HPC sites and commercial Cloud-based resources. Each of these resource types will require some specialization by the frameworks in order to achieve optimal utilization.

\subsection{Grid sites}

The world-wide grid uses batch job schedulers at each site. In general, sites prefer to schedule many independent jobs. These jobs are given a fixed share of resources (number of CPU threads and total CPU memory) they are allowed to use. The fixed resource sharing is needed as sites usually prefer to be able to schedule multiple jobs onto the same computing node. Typically, the computing nodes will have access to the internet as well as their own local disks. Sites also tend to have a site-wide storage system which can be used by workflows to store the files containing the final results of computation before moving those files to an archival storage site. Some sites have special agreements with experiments and allow those experiments to run permanent services at the site, for example a database-caching layer for conditions storage. For frameworks, this means they need internal mechanisms to restrict the number of CPU cores being utilized at a single time and they could rely upon fast local storage during processing. 

\subsection{HPC sites}
\label{sec:hpc_sites}

Unlike grid sites where a ``job slot'' in the system constitutes a fraction of a compute node, an HPC ``job slot'' is typically a cluster of hundreds or thousands of compute nodes that can be used by the program simultaneously. This cluster of compute nodes is allotted a certain amount of time as a group to be run on the site.  It is the responsibility of the user to ensure that a site's compute nodes are well-utilized as poor utilization can negatively impact future resource requests. This translates into a requirement that a large number of nodes must finish work at approximately the same time, which is difficult to accomplish when the processing unit over which work is parallelized is a file containing thousands of events. Further research and development will thus be needed for extending frameworks to efficiently operate in such an environment. This might entail scheduling work for chunks of events not only across all processing elements of a computing node, but also across all nodes that have been assigned to that job. A further challenge is to keep track of the progress of many independent applications processing small chunks of events, where the applications have the ability to stop work after any chunk has finished or use some other mechanism to deal with early termination. One way to achieve this goal is to integrate the application framework with a workflow management system as demonstrated by the Raythena/EventService project~\cite{raythena}. 

In addition, HPC sites vary widely, not only in their computational hardware but also in their network access and storage systems. Many HPC sites do not allow direct internet connections from their worker nodes. Some of these sites do allow internet connections on special edge nodes which, in turn, communicate with the worker nodes. These restrictions can require additional development in order to handle access to resources that are normally shared across computing sites, such as conditions databases. For storage, sites might have local storage on the compute nodes, such as hard disks, SSDs, or maybe even RAM disk, or might not have any local storage. The site-wide storage systems of an HPC site may not be designed to handle small writes from large numbers of processes and, instead, be designed for large writes from a very small number of applications. Further research is likely needed to be able to use the storage on HPC sites effectively.

\subsection{Cloud computing}

Commercial cloud computing offers new possibilities for HEP computing. Experiments could use these resources either to handle a large time-critical data-processing need (e.g. Monte Carlo generation for an upcoming conference) or to supplement their standard computing resources at times when the commercial cloud resources are less expensive. The availability of a large number of compute nodes on short notice could allow experiments to provision their standard computing resources for steady-state processing instead of for peak processing, thereby reducing the total number of compute resources they must maintain. 

Utilizing cloud-computing sites brings a new set of requirements for frameworks. Cloud-computing sites are likely to be used on an opportunistic basis. If the job is being run in a ``spot market'' and will only run for as long as the price to use the CPU is below some threshold, this may require that either the job must terminate very quickly and preserve all work done so far, or the job must periodically store its results into completed files so that intermediate results can be used for later processing. One such approach is to distribute the events of a production job across many semi-independent application workers, each of which process a small chunk of events as described above in section~\ref{sec:hpc_sites}. This allows the preemption of a production job at any time, losing only the work done on the last few events being processed. Correctly handling a job that has only partially completed will likely require additional development of the workflow management system, which is responsible for distributing the jobs across processing sites.

\subsection{Further comments}

Experiments have observed penalties when transferring data into and out of cloud and HPC sites. Addressing this issue could require running all intermediate processing steps for a complete workflow on a single site, rather than distributing results from different processing steps to different computing sites for later processing.

The extreme heterogeneity of computational platforms described in sections~\ref{sec:arch} and~\ref{sec:accelerators}, together with the variety of site execution policies, favoring (e.g.) large-scale, distributed applications on HPC systems, and preemptible applications on commercial clouds, present significant new challenges for HEP computing. Developing data-processing framework modules that run efficiently on accelerators, ideally across multiple platforms, is an effort being pursued by the HEP community in collaboration with HPC experts~\cite{CCE}. Assuming a sufficient number of such modules is available, the challenge for framework developers is to match the resources needed by a module with those provided by one or more computing sites. A further challenge for the framework scheduler is to minimize and ``hide'' data transmission latencies from the memory of one processing element to another. Similar problems have been addressed by distributed application frameworks like Ray~\cite{Ray} and microservice architectures like Dapr~\cite{dapr}, albeit with different performance requirements.

\section{Conclusion}

The recent changes to the computing landscape over the last decade have the potential to meet the computational needs for HEP experiments in the decades to come. To realize that potential, the HEP community will need to continue investing in the development of its data-processing frameworks as fundamental building blocks for exploiting the available computational resources.


\bibliographystyle{JHEP}
\bibliography{references}

\end{document}